%% file: main.tex
\documentclass[conference]{IEEEtran}
\IEEEoverridecommandlockouts

\usepackage{cite}
\usepackage{amsmath,amssymb,amsfonts}
\usepackage{algorithm}
\usepackage{algpseudocode}
\usepackage{textcomp}

\usepackage{mathtools}

\usepackage{tabularx, booktabs}
\usepackage{multirow, makecell}
\usepackage{diagbox}

\usepackage{graphicx}
\graphicspath{{./Figures/}}

\usepackage{epstopdf}
\usepackage{subfigure}
\usepackage{array}
\usepackage{enumerate}
\usepackage{enumitem}
\usepackage{color}

\usepackage{tikz, pgfplots}
\usetikzlibrary{positioning,fit,calc,backgrounds}

\input{preamble.tex}

\definecolor{Blue1}{rgb}{0.3,0,1}

\begin{document}
%

\title{Self-Supervised Learning for Soft-Constrained AC-OPF in Multiphase Distribution Systems
}
\author{Hoang T. Nguyen$^1$, Deepjyoti Deka$^1$, Priya L. Donti$^1$, \\Varsha Pendyala$^2$, Nurali Virani$^2$, Wajid Muneer$^2$, and Reetam Sen Biswas$^2$%
\thanks{ 
	$^1$ Massachusetts Institute of Technology, Cambridge, MA, USA.} 
\thanks{$^2$ GE Vernova Advanced Research Center, Niskayuna, NY, USA.}
}


{}
\maketitle

\begin{abstract}
This paper presents a self-supervised learning framework for fast, soft-constrained AC optimal power flow (OPF) in unbalanced multiphase distribution systems. 
The soft AC-OPF objective balances operating cost, renewable energy curtailment, and penalties for violating operating limits.
A neural network predicts distributed energy resource (DER) setpoints conditioned on operating conditions, switch configurations, and objective weights. The predicted setpoints are then processed through a DER projection and differentiable AC power flow to enforce device limits and network equalities. The framework is trained by directly minimizing the AC-OPF objective, without optimal-solution labels.
Experiments on 123-, 240-, and 906-bus systems yield mean objective gaps of 0.32\%, 0.92\%, and 0.56\%, respectively, relative to IPOPT, across wide operating ranges. 
Across these systems and the evaluated high-load/low-generation and low-load/high-generation scenarios, the largest mean gap is 1.04\%, with mean speedups of 58--5272$\times$ relative to IPOPT.
\footnote{The code is available at \url{https://github.com/MOSSLab-MIT/ACOPF-Multiphase-Distribution-topo-soft}}
\end{abstract}


\section{Introduction}

As distribution networks incorporate a growing proportion of distributed energy resources, system operators need fast AC optimal power flow (AC-OPF) solutions due to the inherent variability of DERs and the need to assess a large number of operating scenarios.
Given the sheer number of DERs and the nonconvex, nonlinear nature of the AC-OPF problem, traditional optimization solvers are simply too slow to support rapid operational decisions and analysis.
In addition, many practical scenarios require soft-constrained formulations that penalize constraint violations rather than enforcing them as hard constraints to handle infeasible operating conditions. In this context, a fast learning-based surrogate offers a promising approach to approximate AC-OPF solutions.

Supervised approaches such as DeepOPF \cite{pan2022deepopf} predict a subset of the decision variables, reconstruct the remaining ones through the AC power-flow equations, and penalize inequality violations during training, but rely on solver-generated optimal solutions as labels.
Recent work on foundation models for transmission-grid analysis includes GridSFM \cite{yang2026gridsfm}, which predicts AC-OPF solutions across diverse grid topologies, and GENCO \cite{puech2026genco}, which provides a unified neural architecture for power flow, OPF, and state estimation. Both combine supervision from solver-generated solutions with physics-based training losses, retaining a dependence on labeled data.

Self-supervised learning approximates AC-OPF solutions by training directly on the objective function and constraints, bypassing the cost of generating optimal solution labels.
DC3 \cite{dontidc3} combines a completion of partial predictions to satisfy equality constraints with gradient-based correction to reduce inequality violations.
FSNet \cite{nguyen2025fsnet} incorporates a differentiable feasibility-seeking step that minimizes constraint violations, enabling end-to-end training.
An augmented-Lagrangian objective is used to train a model that predicts a subset of decision variables and determines the remaining ones through AC power flow \cite{chen2022unsupervised}.
Primal-dual learning jointly trains primal and dual neural networks to approximate primal solutions and instance-specific Lagrange multipliers \cite{park2023self}.


In distribution grids, augmented-Lagrangian-guided learning incorporates instance-specific multipliers and penalties into neural-network training to enable voltage regulation and loss reduction \cite{bouchkati2024augmented}.
Large multiphase distribution feeders additionally require accounting for phase imbalance and switch-induced topology changes.
Our previous work \cite{nguyen2026scalable} addresses these requirements through a self-supervised framework combining differentiable power flow, efficient admittance-inverse updates, and feasibility seeking, with evaluation on unbalanced feeders ranging from 13 to 8,500 nodes.
However, the constraint-handling mechanisms in all of these methods, including our own, aim to satisfy prescribed operating limits as hard constraints; none explicitly prices operating-limit exceedances in a soft-constrained OPF objective.

Compared to our previous work \cite{nguyen2026scalable} on hard-constrained AC-OPF, this paper extends the self-supervised learning approach to a soft-constrained AC-OPF formulation for multiphase distribution systems. 
Voltage magnitude, branch current, angle difference, and substation apparent-power limits are penalized in the objective, while the AC power-flow equations and DER operating limits remain hard constraints. 
This formulation can accommodate challenging operating scenarios in which the prescribed network limits cannot all be satisfied, enabling operators to assess the trade-offs between operating cost, renewable
curtailment, and constraint violations. 
Experiments look at a high level of renewable penetration under a broad range of operating conditions, including high-load/low-DER and low-load/high-DER scenarios. 
Beyond objective gaps and computational speed, we compare individual objective components and the resulting voltage profiles and DER dispatches with those obtained using a nonlinear optimization solver.

\section{Soft-constrained AC-OPF Formulation}
We consider a multiphase distribution feeder with non-slack phase nodes $\cP=\{1,\ldots,N_p\}$ and fixed substation (slack) voltage $\vv_0$ on phases $\Phi_0$.
Let $\cE$ and $\cE_{\mathrm{sw}}$ denote the static multiphase branches and switch branches, respectively.
The prescribed switch-status vector is $\vz=(z_e)_{e\in\cE_{\mathrm{sw}}}$, where $z_e=1$ indicates a closed switch and $z_e=0$ an open switch.
Note that topology is an exogenous input to the AC-OPF problem; we optimize DER dispatch for each operating scenario.
Bold symbols denote vectors or matrices, $\diag(\cdot)$ forms a diagonal matrix, and $\jmath^2=-1$.
An overbar denotes complex conjugation, while $\Re\{\cdot\}$ and $\Im\{\cdot\}$ extract real and imaginary parts.
The positive-part operator is $[x]_+=\max(x,0)$.

The nodal admittance matrix is partitioned as \cite{bazrafshan2017comprehensive}
\begin{equation*}
    \vY(\vz) = \begin{bmatrix}
        \vY_{LL}(\vz) & \vY_{L0}(\vz) \\
        \vY_{0L}(\vz) & \vY_{00}(\vz)
    \end{bmatrix},
\end{equation*}
where subscripts $L$ and $0$ identify non-slack and slack blocks.

Let $\cD$ be the set of DER units, with phase controls indexed by $k=1,\ldots,N_{\text{c}}$.
The set $\cK_j$ contains the phase controls of unit $j\in\cD$.
Active and reactive DERs setpoints form $\vp^{\text{der}},\vq^{\text{der}}\in\R^{N_{\text{c}}}$, with complex vector $\vs^{\text{der}}=\vp^{\text{der}}+\jmath\vq^{\text{der}}$.

For branch $\ell\in\cE$, let $\Phi_\ell$ be its phase set and define the branch--phase pairs $\cE_{\mathrm{ph}}=\{(\ell,\phi):\ell\in\cE,\ \phi\in\Phi_\ell\}$.
For a branch $\ell$, $\vV_{f,\ell}$ and $\vV_{t,\ell}$ are vectors containing its per-unit terminal voltages on the system.
The series-admittance matrix $\vY_\ell^{\text{ser}}$ includes mutual coupling between branch phases.
The branch-current vector $\vi_\ell=(i_{\ell,\phi})_{\phi\in\Phi_\ell}$ is
\begin{equation}
    \vi_\ell=\vY_\ell^{\text{ser}}(\vV_{f,\ell}-\vV_{t,\ell}),
    \quad\ell\in\cE. \label{eq:i_def}
\end{equation}
Let $v_n^{\min},v_n^{\max}$ be voltage-magnitude bounds, $i_\ell^{\text{max}}$ the branch-current rating, and $s_{0,\phi}^{\text{max}}$ the substation apparent-power rating on phase $\phi$.
For the phase endpoints $(m,n)$ of $(\ell,\phi)$, let $t_{\ell,\phi}$ be the tangent of the prescribed angle-difference bound and
\begin{align*}
    d_{\ell,\phi} &= \Re\{\overline{v}_m v_n\}, &
    c_{\ell,\phi} &= \Im\{\overline{v}_m v_n\}.
\end{align*}

Define the voltage, branch-current, angle, and substation capacity penalty functions as
\begin{align}
    P_V=& \sum_{n\in\cP}\Bigl( \relu{(v_n^{\min})^2-|v_n|^2} + \relu{|v_n|^2-(v_n^{\max})^2}\Bigr), \hspace*{-1.1pt} \label{eq:hinge_v}\\
    P_I=&\sum_{(\ell,\phi)\in\cE_{\mathrm{ph}}}
        \relu{|i_{\ell,\phi}|^2-(i_\ell^{\text{max}})^2},\label{eq:hinge_i}\\
    P_\theta =& \sum_{(\ell,\phi)\in\cE_{\mathrm{ph}}}\Bigl(
        \relu{c_{\ell,\phi}-t_{\ell,\phi}d_{\ell,\phi}} + \relu{-c_{\ell,\phi}-t_{\ell,\phi}d_{\ell,\phi}}\Bigr),
        \label{eq:hinge_theta}\\
    P_{\text{sub}} =& \sum_{\phi\in\Phi_0}
        \relu{|s_{0,\phi}|^2-(s_{0,\phi}^{\text{max}})^2}.
        \label{eq:hinge_sub}
\end{align}

In terms of economic efficiency, the operating cost $C_1$ of the system is evaluated as the sum of electricity cost from the substation and DERs:
\begin{align*}
    C_1&=c_1p_0^2+c_2p_0 +\sum_{k=1}^{N_{\text{c}}} \bigl(c_{1,k}^{\text{der}}(p_k^{\text{der}})^2 + c_{2,k}^{\text{der}}p_k^{\text{der}}\bigr),
\end{align*}
where $p_0=\sum_{\phi\in\Phi_0}\Re\{s_{0,\phi}\}$ is the total substation active power.
The substation injections $\vs_0$ are determined by $\vv$ through \eqref{eq:vs0_mat}.
The pairs $(c_1,c_2)$ and $(c_{1,k}^{\text{der}},c_{2,k}^{\text{der}})$ are the quadratic and linear cost coefficients for the substation and DER phase $k$, respectively.
The curtailment cost $C_2$ is 
\begin{equation*}\label{eq:curtailment_cost}
    C_2=\sum_{j\in\cD}\Bigl(p_j^{\text{der,avail}} - \sum_{k\in\cK_j}p_k^{\text{der}}\Bigr).
\end{equation*}

With economic weights $w_{\text{cost}},w_{\text{curtail}}\geq0$ and soft-limit weights $\rho_V,\rho_I,\rho_\theta,\rho_{\text{sub}}>0$, the complete objective is
\begin{equation}
\begin{aligned}
    F&=w_{\text{cost}} C_1+w_{\text{curtail}} C_2\\
    &\quad+\rho_V P_V+\rho_I P_I+\rho_\theta P_\theta+\rho_{\text{sub}} P_{\text{sub}}.
\end{aligned}\label{eq:soft_objective}
\end{equation}
where $p_j^{\text{der,avail}}$ is the available active power of the DER unit $j$.
All objective terms are evaluated at the operating point of the given scenario, and the weights can change across scenarios.

For DER phase control $k$, $p_k^{\text{der,max}}$ and $q_k^{\text{der,min/max}}$ are active- and reactive-power bounds, and $s_k^{\text{der,max}}$ is its apparent-power rating.
The soft-constrained AC-OPF is formulated as
\begin{subequations}\label{prob:ac-opf}
\begin{align}
    \min_{\vp^{\text{der}},\vq^{\text{der}},\vv}
    & F \label{eq:obj}\\
    \st\quad
    &\vi=\vY_{L0}(\vz)\vv_0+\vY_{LL}(\vz)\vv, \label{eq:vi_mat}\\
    &\vs^{\text{der}}-\vs^{\text{load}} =\diag(\vv)\overline{\vi}, \label{eq:power_balance}\\
    &\vs_0=\diag(\vv_0)\overline{\vY_{00}(\vz)\vv_0+\vY_{0L}(\vz)\vv}, \label{eq:vs0_mat}\\
    &\vzero\leq\vp^{\text{der}}\leq\vp^{\text{der,max}}, \label{eq:p_der_lim}\\
    &\vq^{\text{der,min}}\leq\vq^{\text{der}}\leq\vq^{\text{der,max}}, \label{eq:q_der_lim}\\
    &\sum_{k\in\cK_j}p_k^{\text{der}}\leq p_j^{\text{der,avail}},
        \; j\in\cD, \label{eq:p_der_avail}\\
    &(p_k^{\text{der}})^2+(q_k^{\text{der}})^2
        \leq(s_k^{\text{der,max}})^2,\; k=1,\ldots,N_{\text{c}}. \label{eq:s_der_lim}
\end{align}
\end{subequations}

Constraints \eqref{eq:vi_mat}--\eqref{eq:power_balance} impose the network current relation and nodal complex-power balance.
Constraint \eqref{eq:vs0_mat} defines substation injections, and \eqref{eq:p_der_lim}--\eqref{eq:s_der_lim} impose DER device and aggregate availability limits.

\section{Proposed Method}

\subsection{Pipeline}\label{sec:method_overview}
\begin{figure}
	\centering
	\includegraphics[width=0.98\linewidth]{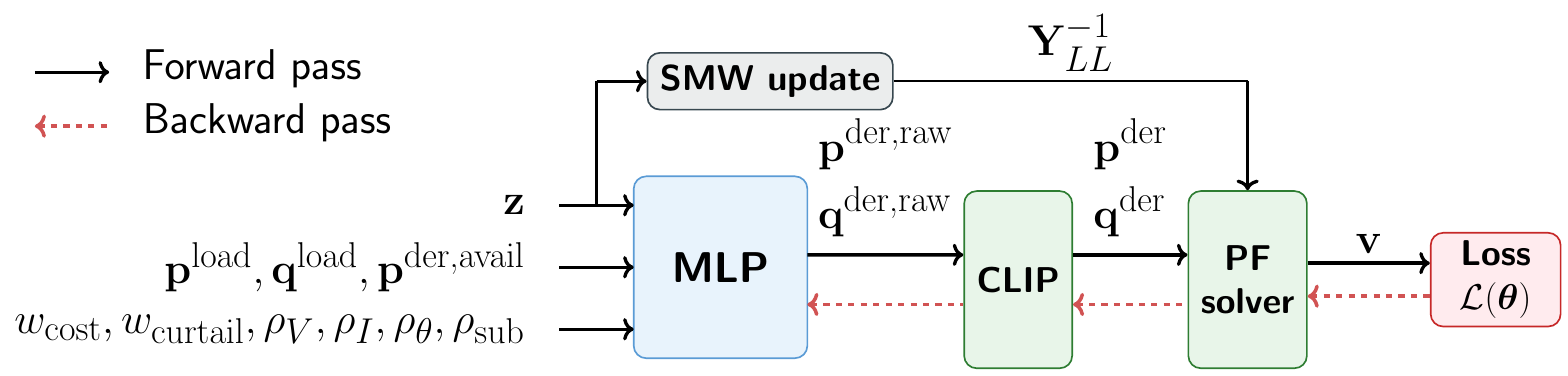}
	\caption{Proposed method for learning to solve soft-constrained AC-OPF.}
	\label{fig:proposed_method}
    \vspace{-0.2in}
\end{figure}

We present the proposed method for training a neural network to solve Problem \eqref{prob:ac-opf} in Fig.~\ref{fig:proposed_method}.
The multi-layer perceptron (MLP) network learns a mapping from the input (loads, DER availability, switch configuration $\vz$, and objective weights) to the output (raw DER setpoints $(\vp^{\text{der,raw}}, \vq^{\text{der,raw}})$).
A clipping layer projects the raw DER setpoints onto the DER operational limits \eqref{eq:p_der_lim}--\eqref{eq:s_der_lim} to obtain $(\vp^{\text{der}},\vq^{\text{der}})$.
A differentiable fixed-point power-flow solver then computes the corresponding voltages $\vv$ needed to evaluate the AC-OPF objective.
When switches change, we use the Sherman-Morrison-Woodbury (SMW) formula to efficiently update the inverse of the admittance matrix $\vY_{LL}(\vz)$, rather than recomputing it from scratch.
For the details of the clip layer, power flow solver, and SMW implementation, please refer to our previous work \cite{nguyen2026scalable}.
Note that the entire process is trained end-to-end and does not require labeled optimal solutions from a traditional AC-OPF solver; instead, it relies directly on the objective and constraint functions to evaluate the solution and update the network parameters to self-improve the predicted solution during training.
During inference, we follow the same pipeline, but without backpropagation or loss computation; instead, we output the complete solution, including $\vp^\mathrm{der}, \vq^\mathrm{der}$, and $\vv$.

\subsection{Loss function}\label{sec:loss}
Let $\vtheta$ denote the neural-network parameters and $\cT$ a training dataset of operating scenarios.
For each scenario $b\in\cT$, $F_b(\vtheta)$ denotes the soft-OPF objective in \eqref{eq:soft_objective}, evaluated at the DER setpoints obtained by projecting the neural-network predictions and the corresponding AC power-flow voltages.
Its dependence on $\vtheta$ arises through the neural-network predictions, the projection layer, and the power-flow computation.
The training loss is the mean objective over a mini-batch $\mathcal{B}$,
\begin{equation}
    \cL(\vtheta)
    =\frac{1}{|\cT|}\sum_{b\in\cT} F_b(\vtheta).
    \label{eq:training_loss}
\end{equation}

The DER limits \eqref{eq:p_der_lim}--\eqref{eq:s_der_lim} are enforced by the projection layer, while the power-flow solver enforces the network equalities \eqref{eq:vi_mat}, \eqref{eq:power_balance}, subject to convergence and numerical tolerance.
Thus, no additional DER-violation or power-flow equality penalty is included in \eqref{eq:training_loss}.
Voltage, current, angle, and substation capacity exceedances are already penalized within $F_b(\vtheta)$ through \eqref{eq:hinge_v}--\eqref{eq:hinge_sub}.
Gradients of \eqref{eq:training_loss} are propagated through the projection and differentiable power-flow solver to update $\vtheta$, i.e., training is conducted end-to-end.

\section{Experimental Results}
We evaluate the proposed method on the IEEE 123-bus, 240-bus, and 906-bus cases, which contain 50, 100, and 61 PV units and 8, 9, and 8 switches, respectively.
Load demands and PV availability are sampled as fractions of the base values in the OpenDSS files under three scenario types.
In the \emph{wide} scenarios, the load and PV availability factors are drawn from $\mathcal{U}[0.2, 1.5]$ and $\mathcal{U}[0, 1]$, respectively, covering the full operating range.
The two \emph{extreme} scenarios restrict sampling to the stressed corners of this range: the high-load/low-DER scenarios use $\mathcal{U}[0.7, 1.0]$ and $\mathcal{U}[0, 0.3]$, and the low-load/high-DER scenarios use $\mathcal{U}[0.2, 0.5]$ and $\mathcal{U}[0.6, 1]$.
For each grid, we generate 200,000 unlabeled training samples, of which 60\% are wide, 20\% high-load/low-DER, and 20\% low-load/high-DER scenarios, and three labeled test sets of 2,000 samples each, one per scenario type.
For the power flow solver, we use fixed-point iteration for the IEEE 123-bus and 906-bus cases, and the Newton-Raphson method for the 240-bus case \cite{nguyen2026scalable}.

We compare against a Direct Prediction baseline, in which the neural network directly predicts all variables, including the DERs' active and reactive setpoints, voltage magnitudes, and voltage angles, without using a power flow solver. Instead, a penalty term for violations of the power flow equations is added to the loss function. For fair comparison, at test time, we pass the predicted DER setpoints through a power flow solver beforecomputing the metrics.

Regarding metrics, we calculate the optimality gap using $(F^\mathrm{pred} - F^\star) / \max\{|F^\star|, 1\}$, where $F^\mathrm{pred}$ is the predicted objective from the proposed method and $F^\star$ is the optimal objective obtained from IPOPT.
As the hard constraints for power flows and DER limits are enforced by construction, we do not separately report model feasibility results.

\subsection{Test results on different scenarios}
The optimality gap and speedup between the models and IPOPT are reported in Table~\ref{tab:test_results}. 
The Direct Prediction approach fails to learn good solutions and suffers from large optimality gaps in all cases.
In contrast, our proposed method closely approximates the IPOPT reference objective while substantially reducing the computational burden of soft-constrained AC-OPF.
On the 123-bus system, the proposed method achieves a small optimality gap of 0.32\% for wide scenarios, and 0.01\% and 0.13\% for the high-load/low-DER and low-load/high-DER scenarios, respectively. Similarly, it obtains gaps of 0.56\% for the 906-bus case and 0.92\% for the 240-bus case under wide scenarios, with similarly low gaps for the extreme scenarios.
Regarding speedup, the proposed method is, on average, 484$\times$ faster than IPOPT for the 123-bus case, 68$\times$ faster for the 240-bus case, and 4548$\times$ faster for the 906-bus case.
This demonstrates potential for repeated AC-OPF evaluations and time-sensitive operational applications.


\begin{table}[t]
    \centering
    \caption{Relative objective gaps (\%, mean $\pm$ std over 3 seeds) and mean speedups over IPOPT (parentheses) on our different test sets.}
    \label{tab:test_results}
    \setlength{\tabcolsep}{2pt}
    \fontsize{6.5}{7.5}\selectfont
    \begin{tabular}{l|lccc}
    \toprule
    Grid & Method & wide & high-load/low-DER & low-load/high-DER \\
    \midrule
    \multirow{2}{*}{123 bus} & Ours & 0.32\std{0.03} ($\times$408) & 0.01\std{0.00} ($\times$626) & 0.13\std{0.01} ($\times$418) \\
     & Direct Prediction & 300.40\std{68.26} ($\times$333) & 47.62\std{60.49} ($\times$515) & 448.44\std{238.39} ($\times$339) \\
    \midrule
    \multirow{2}{*}{240 bus} & Ours & 0.92\std{0.01} ($\times$79) & 0.92\std{0.03} ($\times$59) & 1.04\std{0.01} ($\times$67) \\
     & Direct Prediction & 74.14\std{4.29} ($\times$76) & 8.23\std{0.47} ($\times$58) & 350.10\std{19.69} ($\times$65) \\
    \midrule
    \multirow{2}{*}{906 bus} & Ours & 0.56\std{0.09} ($\times$5272) & 0.04\std{0.01} ($\times$4466) & 0.03\std{0.00} ($\times$3907) \\
     & Direct Prediction & 218.04\std{59.74} ($\times$4521) & 137.28\std{81.42} ($\times$3863) & 82.59\std{17.39} ($\times$3455) \\
    \bottomrule
\end{tabular}
\vspace{-3pt}
\end{table}


\begin{figure}[t]
    \centering
    \includegraphics[width=0.68\linewidth]{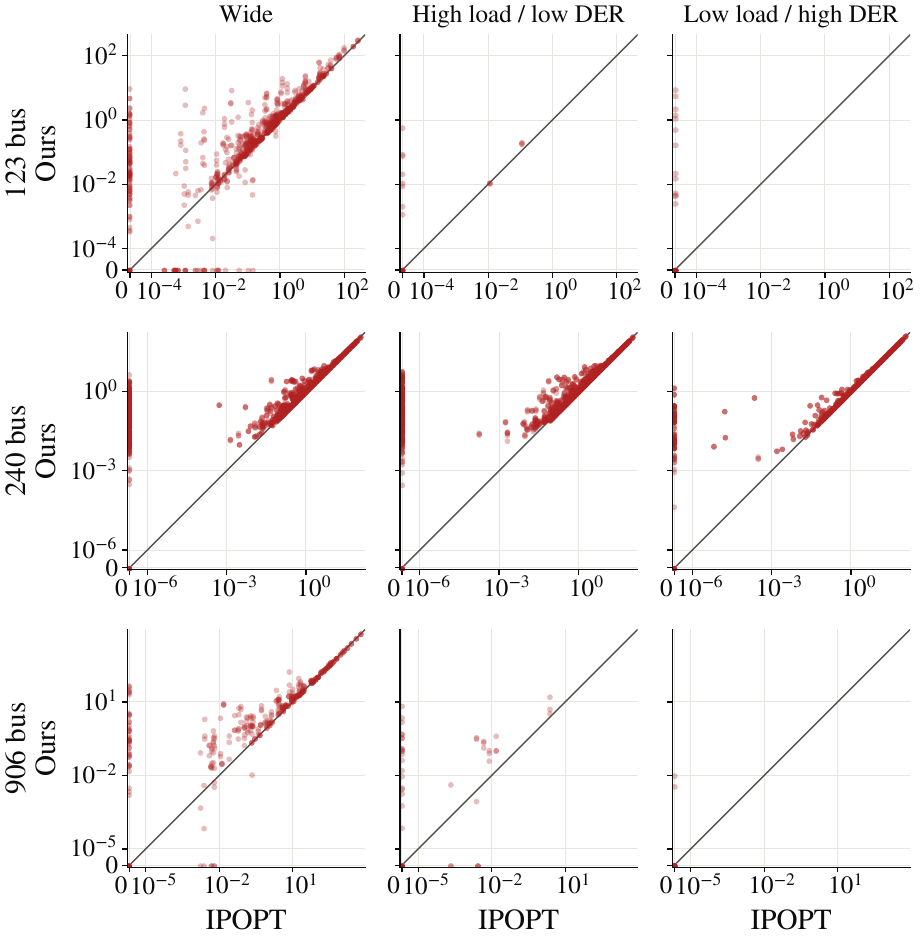}
    \vspace{-4pt}
    \caption{Instance-wise weighted constraint penalties for the proposed method and IPOPT.}
    \label{fig:penalty}
    \vspace{-10pt}
\end{figure}

Beyond the objective gap, we examine how limit violations arise under the two methods. Fig.~\ref{fig:penalty} compares the sum of limit penalties (i.e., $\rho_V P_V+\rho_I P_I+\rho_\theta P_\theta+\rho_{\text{sub}} P_{\text{sub}}$) in \eqref{eq:soft_objective} incurred by our method (vertical axis) and by IPOPT (horizontal axis) instance by instance, pooled over three seeds; points on the diagonal have equal penalties. 
We find that the penalties are driven mainly by voltage constraints and, to a smaller degree, by branch-current constraints; on these test sets, neither method violates the angle-difference or substation constraints.
First, the two methods largely agree on where violations occur. On the wide sets, the proposed method incurs a nonzero penalty in 15.1\%, 92.9\%, and 6.3\% of the scenarios for the 123-, 240-, and 906-bus cases, compared with 14.1\%, 88.2\%, and 5.7\% for IPOPT.
Second, when both methods violate a limit, the penalties are strongly correlated over several orders of magnitude but slightly higher for the proposed method, with mean penalties on the wide sets of 0.91, 6.86, and 11.82, versus 0.69, 6.72, and 11.19 for IPOPT.
Scenarios in which only the proposed method incurs a penalty contribute just 3\%, 13\%, and 6\% of the excess.


\begin{table}[t]
    \centering
    \caption{Relative objective gaps (\%, mean $\pm$ std over 3 seeds) and mean speedups over IPOPT (parentheses) under distribution shift.}
    \label{tab:distribution_shift}
    \setlength{\tabcolsep}{2pt}
    \fontsize{6.5}{7.5}\selectfont
    \begin{tabular}{l|cccc}
    \toprule
    \noalign{\vskip -\belowrulesep}
    \diagbox[height=1.8em, innerleftsep=1pt, innerrightsep=1pt,font=\tiny]{Grid}{Shift} & 0.05 & 0.1 & 0.15 & 0.2 \\
    \noalign{\vskip -\aboverulesep}
    \midrule
    123 bus & 0.25\std{0.04} ($\times$349) & 1.58\std{0.41} ($\times$361) & 5.65\std{0.45} ($\times$362) & 9.50\std{0.54} ($\times$261) \\
    240 bus & 0.76\std{0.00} ($\times$114) & 0.59\std{0.01} ($\times$81) & 0.56\std{0.02} ($\times$83) & 0.49\std{0.02} ($\times$59) \\
    906 bus & 0.64\std{0.04} ($\times$2931) & 2.38\std{0.08} ($\times$3856) & 8.27\std{0.73} ($\times$3606) & 15.80\std{1.22} ($\times$5057) \\
    \bottomrule
    \end{tabular}
    \vspace{-12pt}
\end{table}

Table~\ref{tab:distribution_shift} reports the optimality gaps and speedups under distributional shifts relative to the training data.
For a shift level $\delta$, each load and DER availability in a scenario from the wide test set is multiplied by $1\pm(\delta+\delta|Z|)$, where $Z\sim\mathcal{N}(0,1)$ and the sign is sampled independently for each element. Thus, each perturbation has a relative magnitude of at least $\delta$.
At $\delta=0.05$, the optimality gaps of the proposed method stay at the in-distribution level of Table~\ref{tab:test_results} (0.25\%, 0.76\%, and 0.64\% for the 123-, 240-, and 906-bus cases), i.e., the method is robust to moderate input deviations.

At larger shifts, the mean gaps on the 123- and 906-bus cases grow, reaching 9.50\% and 15.80\% at $\delta=0.2$.
This is expected of a learned surrogate evaluated without retraining, and at $\delta=0.2$, every load and DER availability is perturbed independently by at least 20\%, which increasingly alters the load and DER patterns seen in training.
The degradation is, however, concentrated in a minority of scenarios: at $\delta=0.2$, the median gap is only 0.21\% and 0.06\%, and 74\% and 87\% of the scenarios remain within a 1\% gap.
It also stems almost entirely from voltage penalties, whereas the economic cost stays within about 0.2\% of that of IPOPT.
On the 240-bus case, the relative gap decreases with the shift level because the reference objective roughly doubles (driven by larger voltage penalties in the IPOPT solution) while the absolute objective difference stays nearly unchanged.
The shift experiment tests the trained models, without any adaptation, to isolate their out-of-distribution behavior.
In practice, a natural remedy to even further improve the performance under shifts is to add shifted samples to the training set, which can be done cheaply because training is label-free, i.e., requires no AC-OPF solves.
In addition, as much of the  gap stems from voltage violations rather than economic costs, imposing hard constraints may be beneficial in some cases to maintain robustness under distributional shift, as also observed in our prior work~\cite{nguyen2026scalable}.



\subsection{Histograms}
\begin{figure}[t]
    \centering
    \includegraphics[width=0.95\linewidth]{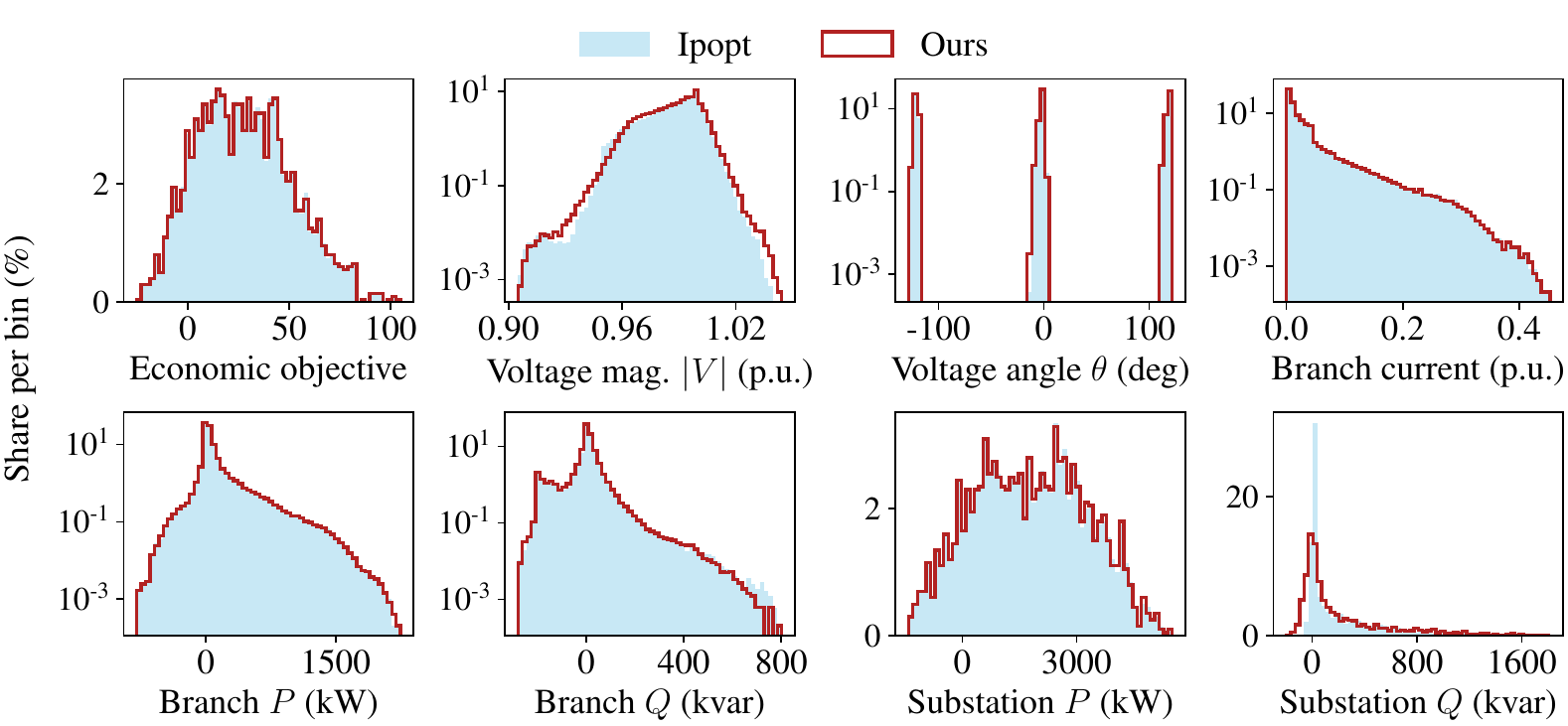}
    \vspace{-4pt}
    \caption{Distributions of the economic objective and electrical quantities over 2,000 wide test instances of the 123-bus system.}
    \label{fig:histogram_123}
    \vspace{-12pt}
\end{figure}

\begin{figure}[t]
    \centering
    \includegraphics[width=0.95\linewidth]{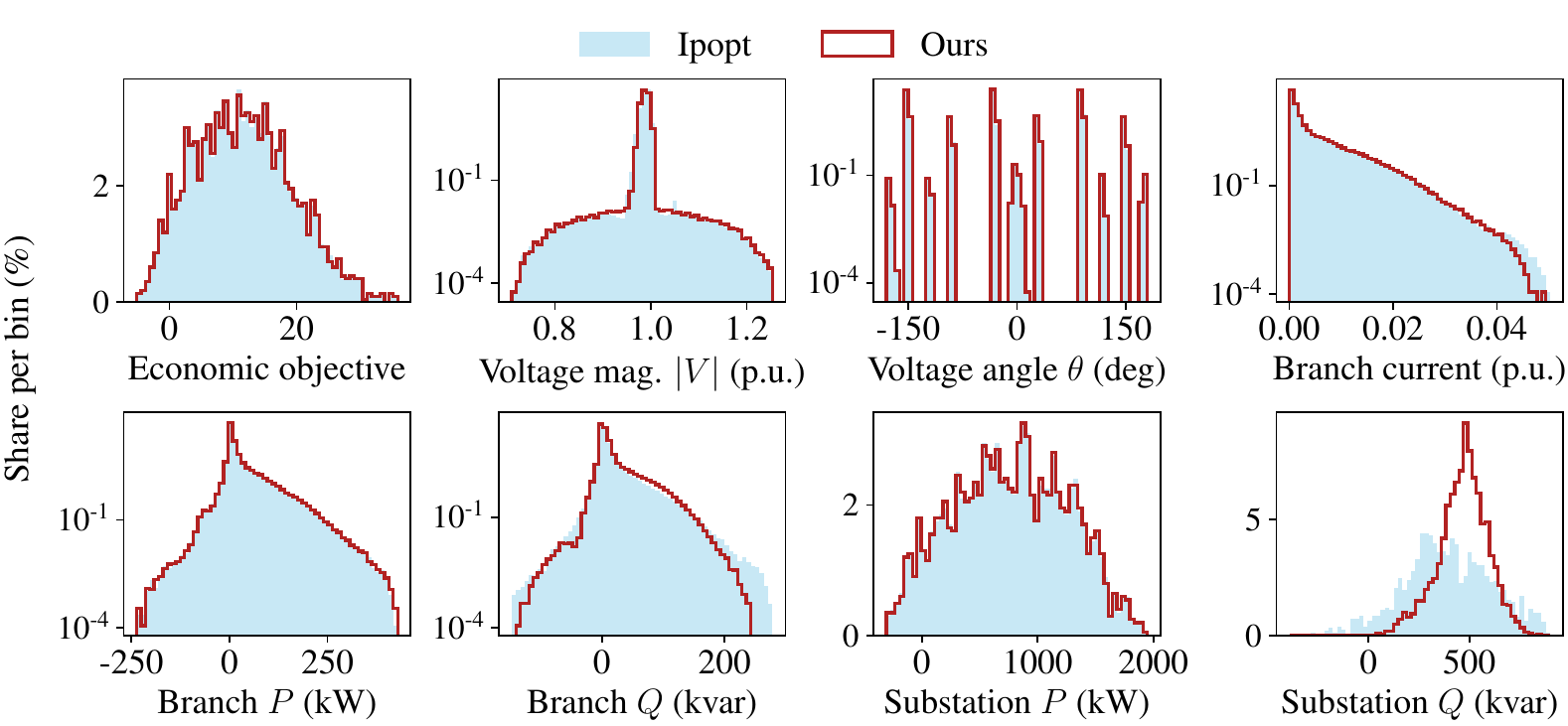}
    \vspace{-4pt}
    \caption{Distributions of the economic objective and electrical quantities over 2,000 wide test instances of the 240-bus system.}
    \label{fig:histogram_240}
    \vspace{-12pt}    
\end{figure}

\begin{figure}[t]
    \centering
    \includegraphics[width=0.95\linewidth]{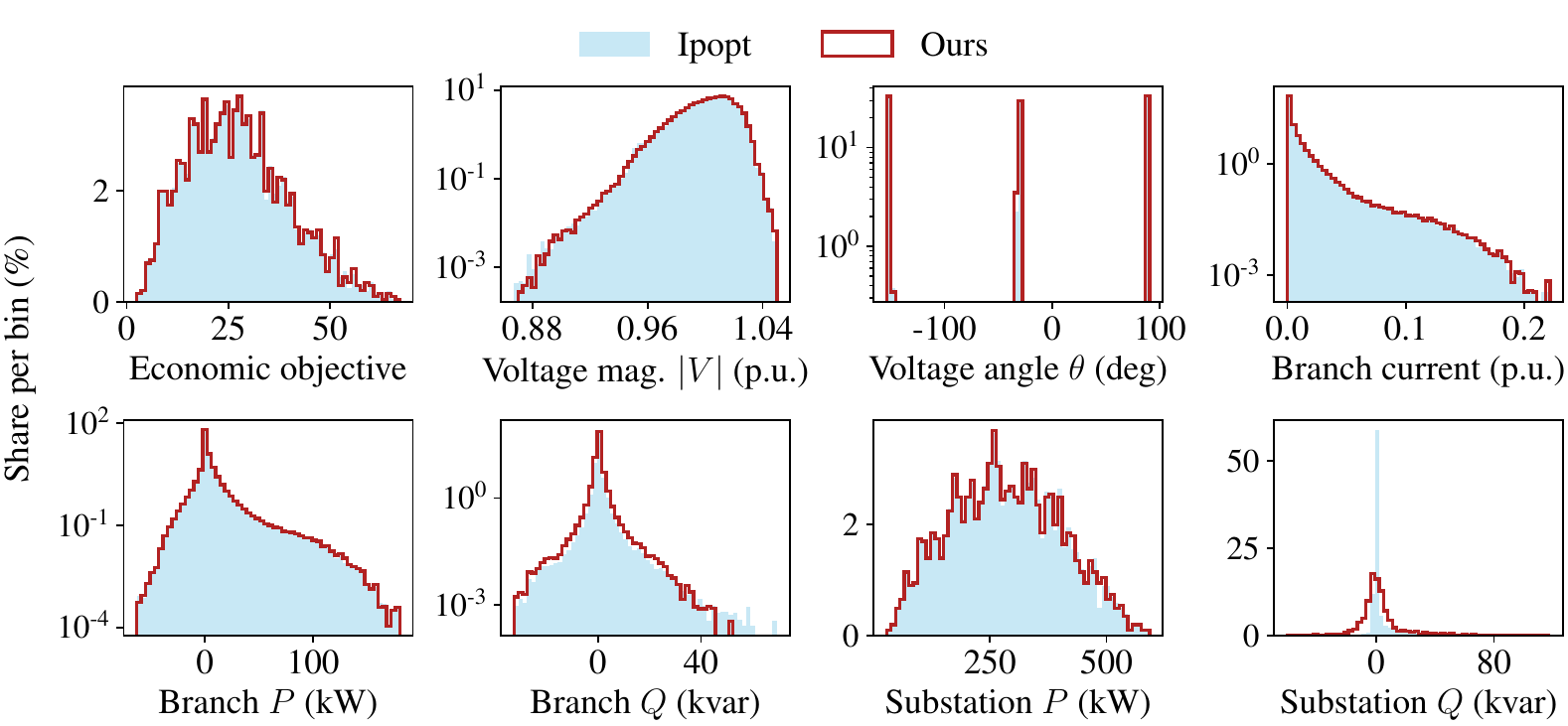}
    \vspace{-4pt}
    \caption{Distributions of the economic objective and electrical quantities over 2,000 wide test instances of the 906-bus system.}
    \label{fig:histogram_906}
    \vspace{-12pt}
\end{figure}

Figs.~\ref{fig:histogram_123}--\ref{fig:histogram_906} 
compare the distributions of the economic objective and key physical quantities for our method and IPOPT over 2,000 wide test scenarios in the 123-, 240-, and 906-bus cases.
Voltage quantities are pooled over non-slack nodes, branch quantities over branch phases, and substation powers 
summed across phases for each scenario.
The distributions of  economic objective, voltage magnitude and angle, branch current, and active power closely overlap across all 
systems.

The most visible differences appear in reactive power.
For the 240-bus case, the proposed method concentrates substation reactive power more tightly around its mean, with a standard deviation of 124~kvar versus 223~kvar for IPOPT.
For the 123-bus case, IPOPT returns substation reactive power near zero in about 30\% of scenarios, whereas 
our method spreads these cases over neighboring values. The 906-bus case shows a similar pattern.
The objective \eqref{eq:soft_objective} does not directly price reactive power; instead, it affects the objective only through active-power losses and the voltage, current, angle-difference, and substation-capacity penalties.
These reactive-power differences are therefore consistent with the small objective gaps reported above, despite the distinct dispatch distributions.

\subsection{Single-instance dispatch and voltage analysis}

\begin{figure}[t]
    \centering
    \subfigure[Voltage magnitude profile]{
        \includegraphics[width=0.46\linewidth]{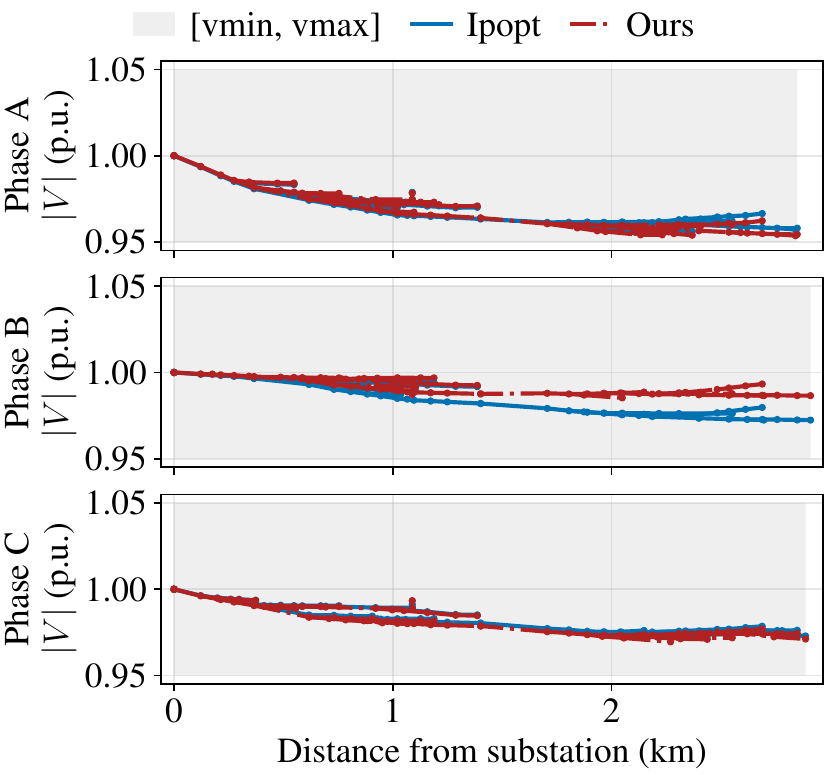}
        \label{fig:3a}
    }
    \hspace{-13.0pt}
    \subfigure[Voltage angle profile]{
        \includegraphics[width=0.46\linewidth]{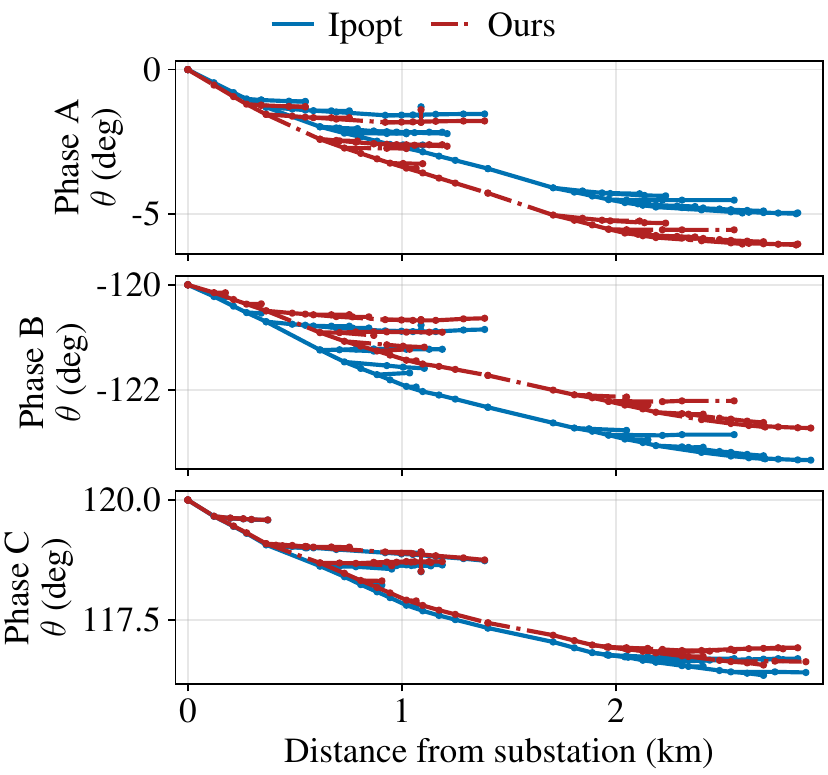}
        \label{fig:3b}
    }
    \vspace{-4pt}
    \caption{Phase-wise voltage magnitudes and angles for our method and IPOPT in a single problem instance of the IEEE 123-bus system.}
    \label{fig:3}
    \vspace{-12pt}
\end{figure}

\begin{figure}[t]
    \centering
    \includegraphics[width=0.90\linewidth]{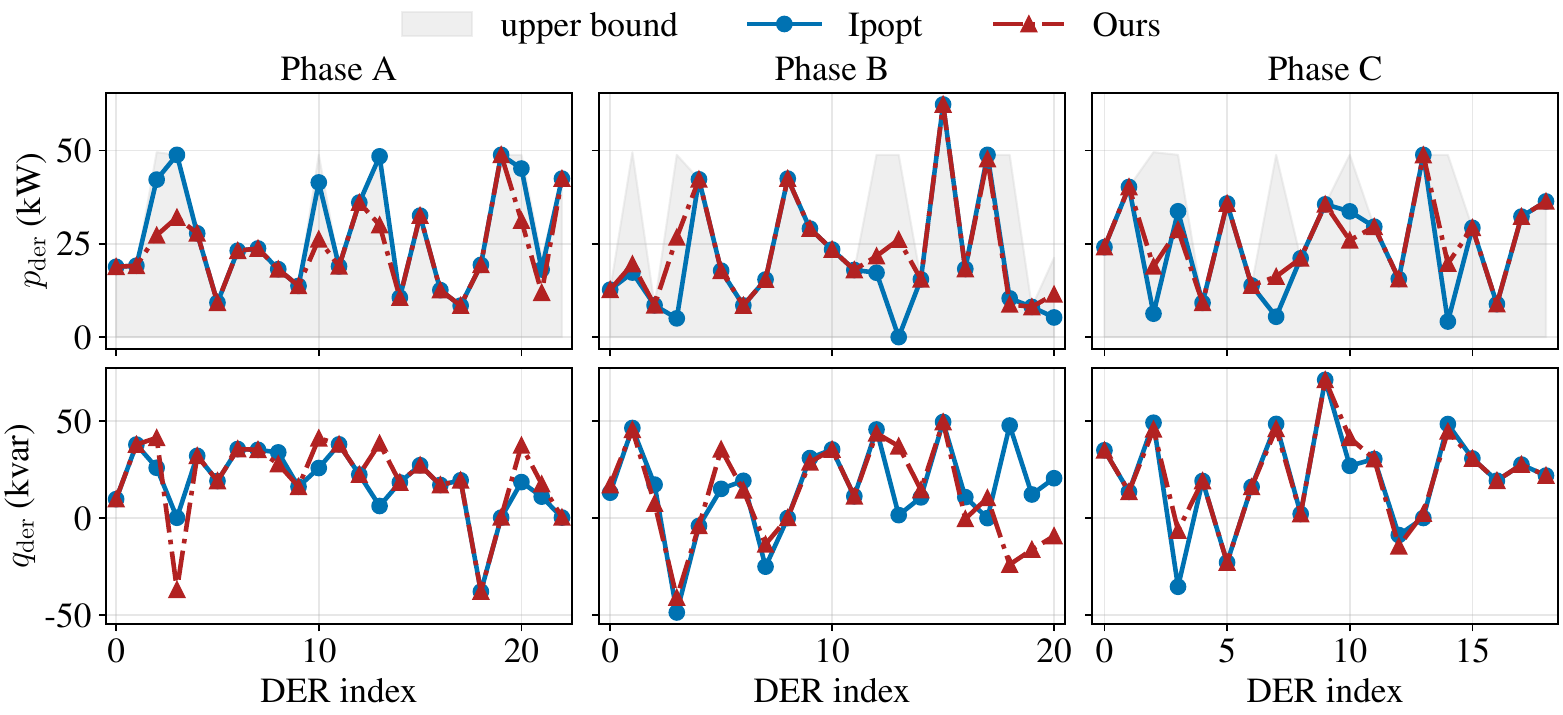}
    \vspace{-4pt}
    \caption{Active and reactive DER setpoints by phase for the proposed method and IPOPT in the same IEEE 123-bus instance as Fig.~\ref{fig:3}.}
    \label{fig:4}
    \vspace{-12pt}
\end{figure}

Figs.~\ref{fig:3} and~\ref{fig:4} show one instance of the 123-bus case in the wide test datasets.
In Fig.~\ref{fig:3a}, the voltage magnitudes produced by the proposed method closely follow those of IPOPT along the feeder on phases A and C.
On phase B, they remain higher toward the feeder end, with a maximum difference of 0.014~p.u., and all magnitudes remain within the limits.
The angles in Fig.~\ref{fig:3b} have the same shape along the feeder and differ by at most $1.1^\circ$.

The DER setpoints in Fig.~\ref{fig:4} agree for most units, but several differ with a maximum difference of 26~kW, and the reactive setpoints differ by up to 72~kvar.
The totals nevertheless match closely.
The proposed method injects 1517.8~kW of the 1519.0~kW available, while IPOPT uses the full 1519.0~kW, and 1127~kvar versus 1117~kvar.
The substation active power in this instance is 2269~kW versus 2265~kW.
The proposed method thus distributes the same total injection differently across the DER phases, changing the objective by only 0.16\%.
This suggests that the objective is flat around the optimum and that a small optimality gap does not require the individual setpoints to match those of IPOPT.

\section{Conclusion}
We presented a self-supervised framework for soft-constrained AC-OPF in unbalanced multiphase distribution systems. Combining DER projection with differentiable AC power flow enables training without optimal-solution labels, while explicit penalties capture trade-offs among operating cost, renewable curtailment, and network-limit violations. 
On 123-, 240-, and 906-bus feeders, the method achieves mean objective gaps of at most 1.04\% relative to IPOPT, with orders-of-magnitude speedups. 
A component-wise comparison shows that the method incurs penalties in nearly the same scenarios as IPOPT, with the remaining gap arising from slightly larger voltage penalties, while the economic cost, voltage profiles, and active-power quantities closely match; individual DER setpoints, particularly reactive power, can differ without materially changing the objective.
Under input shift, the gaps remain near the in-distribution level when loads and DER availability are moderately perturbed, and any degradation is confined to voltage penalties in a minority of scenarios. 
Future work will address this through label-free training-set augmentation, and will extend the framework to real-time optimization and discrete device decisions.


\section*{Acknowledgments}
This work was funded by GE Vernova as part of the MIT-GE Vernova Energy \& Climate Alliance.

\bibliographystyle{IEEEtran}
\bibliography{references}

\clearpage


\end{document}

%% file: preamble.tex
\usepackage{amsmath, amsthm, amsfonts, amssymb}
\usepackage{mathtools}
\usepackage{graphicx}
\graphicspath{{./Figures/}}
\usepackage{subfigure}
\usepackage{booktabs, multirow, array}
\usepackage{bm} 
\usepackage{xcolor}
\usepackage{hyperref}

\definecolor{darkblue}{RGB}{0,0,180}  
\usepackage{tikz}
\usepackage{cancel}
\hypersetup{
  colorlinks=true,     
  linkcolor=darkblue,     
  citecolor=blue,     
  urlcolor=blue        
}

   \def\cD{\mathcal{D}}
\def\cE{\mathcal{E}}   
  \def\cK{\mathcal{K}} \def\cL{\mathcal{L}}
   \def\cP{\mathcal{P}}
   \def\cT{\mathcal{T}}

 \def\vzero{{\bf 0}}

\def\vi{{\bf i}}   
   \def\vp{{\bf p}}
\def\vq{{\bf q}}  \def\vs{{\bf s}} 
 \def\vv{{\bf v}}  
 \def\vz{{\bf z}}

 \def\vV{{\bf V}}  
\def\vY{{\bf Y}} 
\def\vtheta{\bm{\theta}}

\def\R{\mathbb{R}}

\def\st{\text{s.t.}}

\newcolumntype{C}[1]{>{\centering\arraybackslash}m{#1}}

\newcommand{\std}[1]{{\fontsize{4.5}{6}\selectfont \ \textpm {#1}}}

\newtheorem*{theorem*}{Theorem}

\usepackage{thmtools} 

\DeclareMathOperator*{\diag}{diag}

\newcommand{\relu}[1]{\left[ #1 \right]_+}